\documentclass[fleqn,usenatbib,onecolumn]{mnras}
\usepackage[T1]{fontenc}
\usepackage{ae,aecompl}
\usepackage{graphicx}	
\usepackage{amsmath}	
\usepackage{amssymb}	
\usepackage{natbib}
\usepackage{comment}
\newcommand{\VEC}[1]{\vec {#1} }
\title[
Magneto-elastic equilibrium of a neutron-star crust
]
{Magneto-elastic equilibrium of a neutron-star crust}
\author[Y. Kojima, S. Kisaka, K. Fujisawa]{Yasufumi Kojima 
\thanks{%
E-mail: ykojima-phys@hiroshima-u.ac.jp}$^1$,
Shota Kisaka$^1$, 
Kotaro Fujisawa$^2$\\ 
$^1${Department of Physics, 
Graduate School of Advanced Science and Engineering,Hiroshima University,
}\\
{Higashi-Hiroshima, Hiroshima 739-8526, Japan}\\
$^2${Research Center for the Early Universe, Graduate School of Science, University of Tokyo, Bunkyo-ku, Tokyo 113-0033, Japan}
}
\begin{document}
\maketitle 
\begin{abstract}
We examine the equilibrium of a magnetized neutron-star-crust.
We calculate 
 axially symmetric models in which 
an elastic force balances solenoidal motion driven by a Lorentz force.
A large variety of equilibrium models are allowed 
by incorporating the elastic shear deformation; in addition, toroidal-magnetic-field dominated models are available.
These results remarkably differ from those in barotropic fluid stars. 
We demonstrate some models wherein the magnetic energy exceeds the elastic energy. The excess comes from the fact that a large amount of magnetic energy is associated with the irrotational part of the magnetic force, which is balanced with gravity and pressure. It is sufficient for equilibrium models that the minor solenoidal part is balanced by a weak elastic force.
We find that the elasticity in the crust plays an important role on the magnetic-field confinement.
Further, we present the spatial distribution of the shear-stress at the elastic limit, by which the crust-fracture location can be identified.
The result has useful implications for realistic crust-quake models.
\end{abstract}
\begin{keywords}
  stars: neutron -- stars: magnetars -- stars: magnetic fields
\end{keywords}

\section{Introduction}

The magnetic field-strength of neutron stars ranges from  $10^{10}$-$10^{15}$ G at their surfaces. Magnetars possess extremely strong magnetic fields among the species. The interior fields are expected to be considerably stronger than surface fields. A strong magnetic field is also an origin of
various activities, such as bursts and anomalous X-ray fluxes
\citep[e.g.,][for reviews]{2015RPPh...78k6901T,2019RPPh...82j6901E}.
The magnetic-field strength is the largest in the Universe, but the Lorentz force is at best
$\sim 10^{-7}$ times smaller than that of pressure and gravity.
Natural questions arise related to the interior location and configuration of such a strong confined field. 
Theoretical studies that investigate model construction and
the stability of magnetized neutron stars have been performed to deepen our understanding about the same. 
The equilibrium of magnetized stars have been studied in the framework of
Newtonian or relativistic theory of gravity.
A very simple model is a barotropic equilibrium that is contained with a single component, a poloidal or toroidal magnetic field.
However, purely toroidal and poloidal magnetic field configurations are unstable according to an energy principle
\citep{1973MNRAS.161..365T,1973MNRAS.163...77M,1973MNRAS.162..339W}.
Such limiting configurations suffer from
a dynamical pinch-type instability, which the so-called 
the Tayler instability. Further, this instability is explored via numerical MHD simulation
\citep{2004Natur.431..819B,2006A&A...450.1077B,2009MNRAS.397..763B,2011MNRAS.412.1394L,2011MNRAS.412.1730L,2012MNRAS.424..482L,2015MNRAS.447.1213M}.
Dynamical simulations\citep{2009MNRAS.397..763B,2010ApJ...724L..34D} demonstrated
that the final state is a twisted torus configuration, wherein 
 the poloidal and toroidal components of comparable field strengths tangle with each other.
Their stability criterion requires that   
the fraction of the total magnetic energy corresponding to the
toroidal component should exceed 0.2.
Several attempts have been made to calculate static or stationary axially symmetric equilibrium models under various conditions and by using different methods
\citep[e.g.][]{2005MNRAS.359.1117T,2006ApJS..164..156Y,2006ApJ...651..462Y,
2009MNRAS.395.2162L,2010A&A...517A..58D,2013MNRAS.432.1245F,2013MNRAS.434.2480G,
2015ApJ...802..121A,2019PhRvD.100l3019U}.
However, the toroidal magnetic-energy in an equilibrium is not so large.
Instead, some special conditions may be necessary
to construct models with a large toroidal field, for example,
the magnetic field completely confined inside a star \citep{2010A&A...517A..58D},
or a too strong surface-current
\citep{2008MNRAS.385.2080C,2009MNRAS.397..913C,
2013MNRAS.432.1245F,2014MNRAS.445.2777F}.
In addition to the mixed field configuration,
it is also recognized that the stratified structure of density and pressure
is another important factor for stabilizing the system, as previously conjectured
\citep[e.g.,][]{1980MNRAS.191..151T,2012MNRAS.420.1263G,2013MNRAS.433.2445A,2019PhRvD..99h4034Y}.
Recently, \citet{2020MNRAS.499.2636B}
considered the effect of elasticity on magnetic instability.
They performed dynamical simulations of perturbed MHD equations with an elastic force.
Their results show that elastic force suppresses the Tayler instability.
  To save simulation time,
the magnetic-field strength adopted in their paper is rather strong; the maximum is $\sim 10^{17}$G.
 The result is expected to scale down to
 more realistic magnetic-field strengths, e.g., $\sim 10^{14}$G.
Despite some questions that are yet to be answered in the 
general theory of magnetic stars,
there are two important facts 
regarding magnetized neutron stars:
Strong magnetic field $\sim 10^{14}$ G
exists on the surface and 
an elastic force acts on the force balance in the crust below the surface. Therefore, it is quite natural to explore the static equilibrium of the magnetized crust in a neutron star.
Our approach to calculate equilibrium models is complementary to the dynamical approach presented by 
 \cite{2020MNRAS.499.2636B}.
We study the reaction of elastic forces of a crustal ion-lattice to a strong magnetic field.
At first glance, the elastic force appears weak, and therefore, the change in the magnetic energy caused by incorporating elasticity is on the order of the elastic energy in magnitude.
The effect is small for a highly magnetized star.
However, we show that this argument is wrong.
Concrete equilibrium models are crucial for bursts and long-term
magnetic field evolution.
When the elastic deformation exceeds the maximum strain, the crust
may crack and quake.
This event in a magnetar is modeled as a fast radio burst
\citep[e.g.,][]{2019MNRAS.488.5887S,2020ApJ...891...82W,2020ApJ...902L..32D}.
Beyond the critical limit, the ion-lattice may respond plastically.
The consequence of plastic flow on the
Hall evolution of magnetic field is discussed
\citep{2019MNRAS.486.4130L,2020MNRAS.494.3790K,2021MNRAS.502.2097K}, by neglecting the elastic phase.
However, it is unclear which possibility beyond the elastic limit is plausible, and further study of the material property in the crust is required before arriving at a conclusion.
The rest of this manuscript is organized as follows. Section 2 explains models and equations relevant to our problems.
We explore the energy and the structure for various magnetic fields
in equilibrium with the elastic force.
 In Section 3, the numerical results are provided. Finally, our concluding remarks are presented in Section 4.

\section{Model and mathematical formulation}

%
We adopt some simplified assumptions for the crustal model; the density and shear modulus are kept constant.
The elastic deformation of the crust was calculated 
\citep[e.g.,][]{
2000MNRAS.319..902U,2013PhRvD..88d4004J},
to explore the potential sources of continuous gravitational radiation caused by the quadrupole deformation on the rotating neutron star.
The relevant motion is irrotational motion coupled with density perturbation.
We concentrate on a different type of motion, i.e.,
solenoidal elastic-motion without any compression.
For the magnetic field configuration, 
we start with a model wherein the crust is in equilibrium 
without an elastic force.
By changing the magnetic field in the configuration and the magnitude,
the configuration shifts to a new equilibrium with the elastic force.

\subsection{Energy stored in crust}
We estimate the typical magnetic energy stored in a neutron-star crust of radius $R$ and thickness $d$.
The energy $E_{{\rm mag}}$ in the spherical shell for a uniform field with strength $B_{0}$ is given by
\begin{equation}
  E_{{\rm mag}} =\int_{0} ^{2\pi} d\varphi \int_{0} ^{\pi} \sin \theta d\theta 
\int_{R-d} ^{R} r^2 dr \frac{B^2}{8\pi} 
  \approx 
 \frac{1}{20}B_{0}^{2} R^3
 \approx 8.6 \times 10^{44}
 \left(\frac{B_{0} }{10^{14}{\rm G}}  \right)^{2}{\rm erg},
\label{EBestm1}
\end{equation}
where $R = 12{\rm km}$ and $d = 0.1R$.
The maximum elastic-energy $E_{{\rm ela}}$ for a uniform shear modulus ${\bar{\mu}}$ is given by
\begin{equation}
 E_{{\rm ela}} \approx 0.3 \times {\bar{\mu}} \sigma_{c}^2R^3 
 \approx 5.2 \times 10^{43}
 \left(\frac{ {\bar{\mu}}}{10^{30} {\rm erg~ cm^{-3}}}\right)
  \left(\frac{ \sigma_{c}}{10^{-2}}\right)^2{\rm erg} ,
   \label{Eelsestm1}
\end{equation}
where $\sigma_{c}$ is the maximum shear-strain.
The critical value of a neutron star crust
is not well known; it is 
 $\sigma_{c} \approx 10^{-5} -10^{-2}$
 in terrestrial materials\citep[e.g.,][]{1976itss.book.....K};
 however, it is 
 $\sigma_{c} \approx 0.04$ in a semi-analytical approach
  \citep{2018MNRAS.480.5511B} and 
  $\sigma_{c}  \approx 0.1$ in molecular dynamic simulations
 \citep{2009PhRvL.102s1102H,2018PhRvL.121m2701C}.
The shear modulus $\mu$ depends on density 
$\mu \approx 10^{30} {\rm erg /cm^{3}}$
at the core-crust interface
(e.g.,\cite{2008LRR....11...10C}),
and it decreases toward the stellar surface. Therefore, it is unlikely that the elastic energy considerably exceeds the value of eq.(\ref{Eelsestm1}).
On comparing $E_{{\rm ela}}$ with $E_{{\rm mag}}$,
the elastic energy is found to be too small for strongly magnetized neutron-stars.
At a glance, it appears that the elastic force
is ignored to calculate the equilibrium of the solid crust; however, this fact is not true.
Our motivation for this paper is to demonstrate
that elastic force plays an important role in maintaining the equilibrium of the magnetized crust.
The elastic force helps sustain a large amount of magnetic energy $\sim 10^{46}$erg ($\gg E_{\rm ela}$).

\subsection{Formulation}
We consider an equilibrium of the crust in a magnetized neutron star.
We use a spherical coordinate $(r, \theta, \varphi)$ and limit our consideration to an axially symmetric configuration $\partial_{\varphi}=0$.
The static force-balance for a nonrotating star is given by
\begin{equation}
-{\VEC \nabla}P-\rho{\VEC \nabla}\Phi_{\rm G}
+ {\VEC f}+{\VEC h} =0,
  \label{Forcebalance.eqn}
\end{equation}
where the first two terms expressed by pressure $P$, mass density $\rho$, and gravitational potential $\Phi_{\rm G}$ are dominant forces.
In a very good approximation, the structure is spherically symmetric, and  these functions are described by a radial coordinate.
Further, we incorporate the Lorentz force 
${\VEC f} \equiv c^{-1}{\VEC j}\times {\VEC  B}$ and an elastic force ${\VEC h}$ in eq.(\ref{Forcebalance.eqn}).
These forces are considerably smaller than the gravity and pressure force; however, in general, these  cause nonradial acceleration.
The {\it i}~th component $f_{i}$ of the Lorentz force is written as
\begin{equation}
  f_{i}={\nabla}_{j} T^{j}_{i} ,
\end{equation}
where ${\nabla}_{j}$ is a covariant derivative with respect to the three-dimensional metric $g_{ij}$, and 
$T_{ij}$ is a stress tensor for the magnetic field

\begin{equation}
T_{ij} =\frac{1}{4\pi} \left(
B_{i} B_{j}-\frac{1}{2} g_{ij} B^2
\right).
   \label{Temij.eqn}
\end{equation}
The force ${\VEC f}$ is
also written in a vector form as 
\begin{equation}
{\VEC  f}= \frac{1}{c}{\VEC j}\times {\VEC  B}
=\frac{1}{4\pi}\left({\VEC \nabla}\times {\VEC B}\right) \times {\VEC  B}
=-{\VEC \nabla}\left(
\frac{B^{2}}{8\pi} \right)
+\frac{1}{4\pi}\left( 
{\VEC{B}}\cdot {\VEC \nabla}\right) {\VEC  B},
%
\end{equation}
where the electric current ${\VEC j}$ is replaced by the magnetic field by the Amp{\'e}re--Bio-Savart's law.
The Lorentz force ${\VEC f}$
is decomposed as a sum of irrotational and solenoidal vectors, and it is expressed  by a scalar potential $F(r,\theta)$ and 
a vector potential ${\VEC A}(r,\theta)$ 
\begin{equation}
  {\VEC f}=
  -{\VEC{\nabla}}F
  + {\VEC{\nabla}}\times {\VEC A}.
   \label{decmpFA.eqn}
\end{equation}
These functions are obtained by solving the following equations with appropriate boundary conditions
\begin{align}
\nabla^2 F =-\VEC{\nabla}\cdot {\VEC f},
\label{eqndf2}
\\
{\VEC{\nabla}}\times {\VEC{\nabla}}\times {\VEC A}
= \VEC{\nabla}\times{\VEC{f}}.
\label{eqncirc}
\end{align}
The irrotational part of ${\VEC f}$ is closely coupled to the dominant forces.
The potential $F$ may be regarded as
an additional pressure-term in eq.(\ref{Forcebalance.eqn}).
However, the solenoidal part
is almost irrelevant to the dominant forces.
Here, we assume barotropic distribution between
$P$ and $\rho$, i.e., 
$ {\VEC{\nabla}}\times (\rho^{-1} {\nabla}P)=0$, so that
the ``curl'' of the first two terms in eq.(\ref{Forcebalance.eqn}) vanishes.
The solenoidal part of ${\VEC f}$ drives a shear motion. We assume that an elastic force
${\VEC h}$ is balanced with it in the crust.
The elastic force ${\VEC h}$ is also smaller than the dominant forces in magnitude, like the Lorentz force.
There are two approached to vector decomposition with respect to 
force density ${\VEC f}$ or acceleration  $\rho^{-1}{\VEC f}$.
In this paper, we assume $\rho$ is constant in the crust and that they are identical. 
The {\it i}~th component $h_{i}$ is expressed by the elastic displacement vector $\xi_{i}$ and a shear modulus $\mu$ as
\begin{equation}
h_{i}= {\nabla}_{j} \left(\mu \sigma^{j}_{i} \right).
 \label{ElasticForce.eqn}
\end{equation}

Here, we assume incompressibility 
${\nabla}_{k} \xi ^{k}=0$. The shear tensor
$\sigma_{ij}$ is given by\footnote{%
Explicit forms in a spherical coordinate 
are written in some text books
(e.g.,\cite{1959thel.book.....L}).
}
\begin{equation}
\sigma_{ij} ={\nabla}_{i} \xi _{j} +{\nabla}_{j} \xi _{i}.
 \label{sheartensor.eqn}
\end{equation}
For simplicity, we assume that $\mu$ is constant in the entire crust, and 
${\VEC h}$ is expressed in the vector form as
\begin{equation}
{\VEC h}=-\mu 
{\VEC{\nabla}}\times {\VEC{\nabla}}\times  {\VEC \xi} .
\label{Shear.eqn}
\end{equation}
This form clearly shows
that ${\VEC h}$ is a solenoidal vector.
This term is balanced with the solenoidal part of the Lorentz force in eq.(\ref{Forcebalance.eqn}), and
the vector potential ${\VEC A}$ in eq.(\ref{decmpFA.eqn}) is therefore expressed as
\begin{equation}
{\VEC A}=\mu {\VEC{\nabla}}\times  {\VEC \xi} .
\label{abyxi.eqn}
\end{equation}
We explicitly write equations
for the poloidal components
${\VEC \xi}_{p}\equiv(\xi_{r}, \xi_{\theta})$ and the azimuthal component $\xi_{\varphi}$
to determine the magneto-elastic equilibrium.
The azimuthal component 
$({\VEC f}+{\VEC h})_{\varphi}=0$ of eq.(\ref{Forcebalance.eqn}) is reduced to
\begin{equation}
 c\mu  [{\VEC{\nabla}}\times {\VEC{\nabla}}\times 
 (\xi_{\varphi} {\VEC e}_{\varphi} )]_{\varphi} =
 \left( {\VEC j} \times {\VEC B}  \right)_{\varphi}.
   \label{Forcebalance1.eqn}
\end{equation}
The azimuthal component of eq.(\ref{eqncirc})
is reduced to
\begin{equation}
 c\mu \left[ {\VEC{\nabla}}\times {\VEC{\nabla}}\times 
\left(\frac{W}{\varpi} {\VEC e}_{\varphi}\right) \right]_{\varphi}
= \varpi\left[
{\VEC B} \cdot {\VEC \nabla}\left(\frac{j_{\varphi}}{\varpi} \right) 
 -{\VEC j} \cdot {\VEC \nabla}\left(\frac{B_{\varphi}}{\varpi} \right)\right] ,
\label{Forcebalance2.eqn}
\end{equation}
where $\varpi=r\sin\theta$ is the cylindrical radius,
$W\equiv\mu^{-1}A_{\varphi}\varpi$,
and the conditions 
${\VEC \nabla}\cdot{\VEC j}={\VEC \nabla}\cdot{\VEC B}=0$ are used.
Other components in eq.(\ref{eqncirc}) except for the azimuthal component vanish 
by eq.(\ref{Forcebalance1.eqn}) and an axial symmetry ($\partial_{\varphi}=0$).
It is convenient to introduce a scalar functions $X(r,\theta)$ to express ${\VEC \xi}_{p}$ as
\begin{equation}
{\VEC \xi} _{p}= {\VEC{\nabla}}\times \left(\frac{X}{\varpi} {\VEC e}_{\varphi}\right).
  \label{XibyX.eqn}
\end{equation}
The azimuthal component of eq.(\ref{abyxi.eqn}), a relationship between $W(=\mu^{-1}A_{\varphi}\varpi)$ and ${\VEC \xi}_{p}$, is reduced to
\begin{equation}
\left[ {\VEC{\nabla}}\times {\VEC{\nabla}}\times 
\left(
\frac{X}{\varpi} {\VEC e}_{\varphi} \right) \right]_{\varphi} 
=\frac{W}{\varpi}.
\label{Elliptic3.eqn}
\end{equation}

Thus, for given magnetic fields, elastic reactions are determined by solving three  second-order partial differential equations
(\ref{Forcebalance1.eqn}), (\ref{Forcebalance2.eqn}), and (\ref{Elliptic3.eqn}) of an elliptic type.
 Our calculation is limited to a spherical shell-region of the crust of a neutron star,
 $r_{c} \le r \le R$ and $0\le \theta \le \pi$.
The crust thickness $d\equiv R-r_{c}$ is assumed to be $d/R=0.1$.
The exterior region of $r>R$ is treated as vacuum.
We discuss the boundary conditions for these functions, $\xi_{\varphi}$, $W$, and $X$. 
The displacement of the axial component vanishes
$\xi_{\varphi}=0$ at the core-crust interface  $r=r_c$, 
whereas the shear is free
$\sigma_{r\varphi}=0$ at 
$r=R$, which is given by
$\partial_{r}(\xi_{\varphi}/r)=0$.
The solenoidal motion is assumed to be along to radial boundaries, i.e., 
zero displacement $\xi_{r}=0$ and zero
acceleration $({\VEC{\nabla}}\times {\VEC{A}})_{r}=0$.
These conditions
are $W=X=0$ at $r=r_{c}$ and $R$.
Finally, all functions 
$\xi_{\varphi}$, $W$, and $X$ should vanish 
on the symmetric axis ($\theta =0, \pi$) by a regularity condition.

\subsection{Magnetic field in barotropic equilibrium}

An axially symmetric magnetic-field is described by two functions
\begin{equation}
\VEC{B}=\VEC{\nabla}\times \left(
\frac{\Psi}{\varpi}\VEC{e}_{\varphi}\right)
+\frac{S}{\varpi}\VEC{e}_{\varphi} .
\label{eqnDefBB}
\end{equation}
The current function $S$ should be a function of  $\Psi$ for the azimuthal component of the Lorentz force 
to vanish.
The azimuthal current in
 a barotropic MHD equilibrium with an axial symmetry  is written as 
\begin{equation}
\frac{4\pi\varpi}{c}j_{\varphi} = -\rho K^\prime \varpi^2 -S^\prime S,
 \label{MHDeqil.eqn}
\end{equation}
where $K$ is a function of $\Psi$ and  
a prime denotes a derivative with respect to $\Psi$.
The Lorentz force is written as
\begin{equation}
 {\VEC f}=- \frac{\rho K^\prime}{4\pi}{\VEC \nabla}\Psi.
\end{equation}
Thus, there is no 
solenoidal force, i.e., an
 elastic force vanishes, ${\VEC \xi}=0$ for this field.
For the given functional forms $S(\Psi)$ and $K(\Psi)$, 
the magnetic function $\Psi$,
which describes the poloidal field ${\VEC B}_{\rm p}$,
is solved for the source term (\ref{MHDeqil.eqn}).
The method is straightforward; however, the actual models obtained by numerical calculations are limited.
The solution is not obtained for arbitrary forms $S(\Psi)$ and $K(\Psi)$.
This fact suggests that
the barotropic condition may highly constrain static models.
A simple solution in a barotropic MHD-equilibrium is easily obtained 
 by assuming that the field is purely 
dipolar ($\Psi \propto \sin^2\theta$ and $S=0$),
and that $\rho K^\prime$ is a constant.
We adopt this magnetic-field configuration as a fiducial model, and we examine the response of 
the elastic force by adding an extra field.
We consider the two cases for the magnetic field at the core-crust interface $r_{c}$.
These conditions named type-I and type-II model
the magnetic field in the core.
For the type-I case, the magnetic field is expelled, and therefore, 
we impose $\Psi =0$ at $r=r_{c}$.
The azimuthal current inside the crust produces a dipolar field.
The magnetic field penetrates to the core
for the type-II case.
The dipolar field is imposed for $\Psi$ at $r=r_{c}$, and 
there is no current inside the crust\footnote{%
    Magnetic field is expelled or penetrates depending on
    the type of superconductor in the core.
    More sophisticated treatment is necessary to describe
    realistic type-II superconductor
    (see \cite{2013MNRAS.431.2986H}).}.
In both cases, the function $\Psi$ is smoothly connected to the external dipole
field in vacuum.

\subsection{Models
departed from barotropic equilibrium}
 
The magnetic field and electric current in a barotropic equilibrium are expressed
with a subscript ``${(0)}$'' like
 ${\VEC B}_{(0)}$ and ${\VEC j}_{(0)}$, in which ${\VEC \xi}=0$.
We modify the magnetic field and relax the assumption $S(\Psi)$
to examine the equilibrium model
with elastic force (${\VEC \xi}\ne0$).
We consider a combined field
${\VEC B}_{(0)}+{\VEC B}_{({\rm s})} $ and current 
${\VEC j}_{(0)} +{\VEC j}_{({\rm s})}$,
where  subscript  (s = A, B, C) denotes the model described below.
The magnetic flux or electric current is consistently solved by the 
Amp{\'e}re--Bio--Savart's law for a given
${\VEC B}_{({\rm s})}$ or ${\VEC j}_{({\rm s})}$.
The angular function of the extra field is given by the Legendre polynomials
$P_{l} (\cos \theta)$ with a multipole index $l$.
The radial function is a quartic of $r$ in the crust
($r_c\le r \le R$),
and it vanishes both at the core-crust interface and at the surface.
%

  \begin{description}
    \item[Model A] Poloidal magnetic field 
    extended to the external
%
\begin{equation}
~~~\frac{4\pi\varpi}{c} j_{\varphi}=N_{{\rm A}}B_{0}R^{-4}[(r-r_c)(r-R)]^2
\sin{\theta}P_{l}^{\prime},
\label{loopj3}
\end{equation}
    \item[Model B] Poloidal 
    magnetic field confined within the crust
%
\begin{equation}
~~~\Psi=N_{{\rm B}}B_{0}R^{-2}[(r-r_c)(r-R)]^2\sin{\theta}P_{l}^{\prime}.
\label{loopPsi}
\end{equation}
    \item[Model C] Toroidal magnetic field
    confined within the crust
%
\begin{equation}
~~~S= \varpi B_{\varphi}=N_{{\rm C}}B_{0}R^{-3}[(r-r_c)(r-R)]^2 \sin{\theta}P_{l}^{\prime}.
\label{loopB3}
\end{equation}
  \end{description}
In eqs.(\ref{loopj3})--(\ref{loopB3}),
a prime denotes a derivative
 with respect to $\cos \theta$, and 
$N_{{\rm s}}$(s = A, B, C) is a normalization constant.
We use the normalization 
$B_{0}\equiv|B_{0}(R,0)|$ of
the magnetic-field strength at the magnetic pole for the fiducial field.
The boundary condition is type-I or type-II
at the core-crust interface for the fiducial 
magnetic-field, and the 
extra field is imposed by Model A, B or C; therefore, we consider six different models 
listed in Table~\ref{table1:mylabel}. 
Further, we consider a mixed poloidal--toroidal model
simultaneously imposed according to B and C
in Section 3. 

\begin{table}
    \centering
    \begin{tabular}{c|l|l} \hline
    Model     &  Magnetic field geometry & Results \\ \hline 
   AI     & 
     Poloidal field by eq.(\ref{loopj3}) and dipolar field expelled at $r_{c}$  
   &  Fig.~\ref{FigE.ACI}, Fig.~\ref{Figbx6-I}, Fig.~\ref{FigMagtoEl}\\
   BI     &  Confined poloidal field by eq.(\ref{loopPsi}) and dipolar field expelled at $r_{c}$
   &  Fig.~\ref{FigE.ACI}, Fig.~\ref{Figbx6-I}, Fig.~\ref{FigMagtoEl}\\
   CI     &   Confined toroidal field by eq.(\ref{loopB3})  and dipolar field expelled at $r_{c}$ 
   &  Fig.~\ref{FigE.ACI}, Fig.~\ref{Figbx6-I}, Fig.~\ref{FigMagtoEl}\\
   AII        &
   Poloidal field by eq.(\ref{loopj3}) and dipolar field penetrated at $r_{c}$   
   &  Fig.~\ref{FigE.ACII}, Fig.~\ref{Figbx6-II}, Fig.~\ref{FigMagtoEl}\\
   BII     & 
   Confined poloidal field by eq.(\ref{loopPsi}) and dipolar field penetrated at $r_{c}$ 
   &   Fig.~\ref{FigE.ACII}, Fig.~\ref{Figbx6-II}, Fig.~\ref{FigMagtoEl}\\
   CII     &
   Confined toroidal field by eq.(\ref{loopB3}) and dipolar field penetrated at $r_{c}$  
   &  Fig.~\ref{FigE.ACII}, Fig.~\ref{Figbx6-II}, Fig.~\ref{FigMagtoEl}\\
       Mixed     &
       Mixed poloidal--toroidal field by eqs.(\ref{loopPsi}),(\ref{loopB3}) and 
       dipolar field expelled at $r_{c}$ 
       & Fig.~\ref{Fig6}, Fig.~\ref{Fig7}, Fig.~\ref{Fig8}, Fig.~\ref{Fig9}\\ \hline
    \end{tabular}
    \caption{Description of models and relevant results in Section 3}
    \label{table1:mylabel}
\end{table}

   \subsection{Energy}
We provide the expression of energy stored inside a crust of volume.
The magnetic energy $E_{\rm mag}$ is 
given by
\begin{equation}
E_{\rm mag} =\int_{V} \frac{1}{8\pi} B^2 dV.
%
\end{equation}
We divide it as the energy of the poloidal component
$E_{\rm mag,P}$ and the toroidal component $E_{\rm mag,T}$, $E_{\rm mag} =E_{\rm mag,P}+E_{\rm mag,T}$.
Elastic energy in a case of constant shear modulus is obtained by 
\begin{equation}
E_{\rm ela}=\mu \int_{V}  (\VEC{\nabla}\times\xi)^2 dV.
%
\end{equation}
The same result for $E_{\rm ela}$ is also calculated by
the shear tensor because 
\begin{equation}
 \frac{\mu}{2} \sigma_{ij} \sigma^{ij} 
= \mu[(\VEC{\nabla}\times {\VEC{\xi}})^2 -\VEC{\nabla} \cdot
\VEC{\mathcal{V}}_{\rm p}],
%
\end{equation}
where $\VEC{\mathcal{V}}_{\rm p}$ is a poloidal
 vector, whose component is given by
 \begin{equation}
\VEC{\mathcal{V}}_{\rm p}=\left[ \partial_{r} (\xi_{r})^2
+\frac{2\xi_{\theta}}{r}( \partial_{\theta}\xi_{r}-\xi_{\theta}),
~
\frac{1}{r}\partial_{\theta} (\xi_{\theta})^2
+\frac{2\xi_{r}}{\varpi}\partial_{r}(\varpi \xi_{\theta})\right].
%
\end{equation}
The volume integral of the divergence of a vector
$\VEC{\mathcal{V}}_{\rm p}$ is replaced by surface integrals
at $r=r_{c}$ and $R$, which vanish.

\subsection{Elastic limit and parameter}
In the fiducial magnetic-field configuration,
there is no elastic shear-deformation ${\VEC \xi}=0$, and therefore, $E_{\rm ela}=0$.
Equilibrium models with ${\VEC \xi} \ne 0$ are constructed as follows.
We specify the magnetic field
${\VEC B}_{(0)}+{\VEC B}_{({\rm s})}$ and the current 
${\VEC j}_{(0)} +{\VEC j}_{({\rm s})}$ with non-zero value of
$N_{{\rm s}}$ in eqs.(\ref{loopj3})--(\ref{loopB3}).
The functions $\xi_{\varphi}$, $W$, and $X$ are obtained by
numerically solving partial differential equations
(\ref{Forcebalance1.eqn}), (\ref{Forcebalance2.eqn}),
and (\ref{Elliptic3.eqn})
for their source terms calculated by the magnetic field.
The poloidal displacement is also obtained by eq.(\ref{XibyX.eqn}).
The displacement $|\xi|$ and the elastic energy $E_{\rm ela}$ generally increase with the normalization constant $N_{{\rm s}}$ in eqs.(\ref{loopj3})--(\ref{loopB3}).
The maximum value of $|\xi|$ or the shear tensor $|\sigma_{ij}|$ 
(eq.(\ref{sheartensor.eqn})) is limited by
an empirical criterion.
  We adopt the von Mises criterion\citep[e.g.,][]{
    1969imcm.book.....M,2000MNRAS.319..902U}
  to estimate the maximum
\begin{equation}
\frac{1}{2}\sigma_{ij}\sigma^{ij}
\le (\sigma_{c})^{2},
   \label{criterion}
\end{equation}
where $ \sigma_{c}$ is the maximum strain.
  As another criterion, the Tresca one is based on the difference
  between the maximum and minimum of shear.
  Two criteria in general predict different critical states
  \citep[e.g.,][]{1969imcm.book.....M}. 
Our concern is order-of-magnitude level.
More precise discussion requires further improvement of the criterion.
The elastic limit (\ref{criterion}) thus constrains the normalization constant $N_{{\rm s}}$.
From now on, we discuss this critical state for a given magnetic field.
Our basic equations (\ref{Forcebalance1.eqn}) and (\ref{Forcebalance2.eqn}) in the maximum strain contain a dimensionless parameter denoted as $q$
\begin{equation}
 q\equiv\frac{4 \pi \mu \sigma_{c}}{B_{0}^2 }
 \approx 1.3  \times 10^{1}
 \left(\frac{B_{0} }{10^{14}{\rm G}}  \right)^{-2}
 \left(\frac{ \mu}{10^{30} {\rm erg~cm^{-3}}}\right)
  \left(\frac{ \sigma_{c}}{10^{-2}}\right).
%
\end{equation}
Here, $B_{0}$ is magnetic field strength at the pole.
This parameter corresponds to a ratio of the elastic to magnetic forces.
The limit of $q=0$ corresponds to our fiducial model. 

 \section{Numerical Results}
%

   \subsection{Energy of
   magneto-elastic equilibrium model}

\begin{figure}\begin{center}
  \includegraphics[scale=0.85]{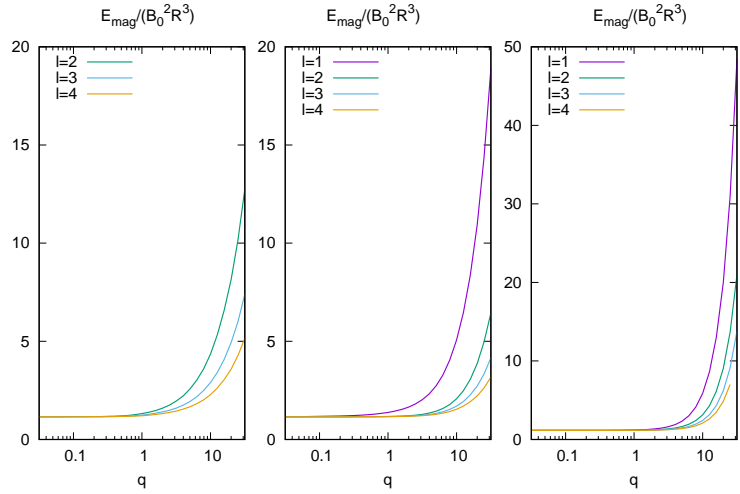}%
\caption{ 
 \label{FigE.ACI}
The magnetic energy $E_{\rm mag}/(B_{0}^2R^3)$ in magneto-elastic equilibrium as a function of $q$. The left to right panels show results for extra components of the multi-pole $l$ in Models A-I, B-I, and C-I.
}
\end{center}\end{figure}
\begin{figure}\begin{center}
  \includegraphics[scale=1.0]{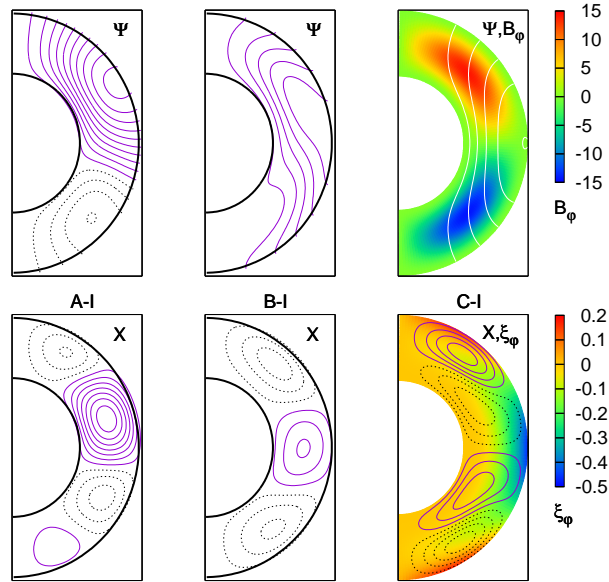}%
\caption{ 
  \label{Figbx6-I}
  Magnetic field in top panels and elastic displacement in bottom panels in ``enlarged'' $r$-$\theta$ coordinate for  Model A-I, Model B-I and Model C-I with $l=2$ and $q=10$.
 Contour intervals of $\Psi$ are $\Delta \Psi/(B_{0}R^2) = 0.1$.
Color contour in the right panel represents toroidal field $B_{\varphi}/B_{0}$.
Contour intervals of $X$ in the bottom panels are $\Delta X/(\sigma_{c} d^3) =0.1$ in the left and middle panels, whereas it is $\Delta X/(\sigma_{c} d^3) =0.01$ in the right panel.
The color contour in the right panel represents the toroidal field $\xi_{\varphi}/(\sigma_{c} d)$.
The dotted lines represent the negative value of the contour level.
}
\end{center}\end{figure}

We assume that the magnetic field is expelled at the inner radius of the crust, and the extra field is imposed  according to eqs.(\ref{loopj3})--(\ref{loopB3}).
These models are labeled as Model A-I, B-I, and C-I (see Table~\ref{table1:mylabel}).
Further, they depend on the strength parameter $q$ and multipole $l$.
We always fix the magnetic dipole field at the surface, and thus, we consider $l\ge 2$ in Model A, but $l\ge 1$ in the other models.
Figure~\ref{FigE.ACI} shows a magnetic energy in an equilibrium with the maximum shear-strain.
In weak elastic-force cases $q \ll 1$,
the magnetic energy does not differ from that of a barotropic equilibrium model, in which the energy is given by
\begin{equation}
 E_{{\rm mag}} ^{(0)} \approx  1.2 B_{0}^2R^3
 \approx 2.1 \times 10^{46} 
 \left(\frac{B_{0} }{10^{14}{\rm G}}  \right)^{2}
  {\rm erg},
 \label{engBI0}
\end{equation}
where the typical values for the radius $R = 12{\rm km}$, and
the crust thickness $d = 0.1R$ is used.
The value (\ref{engBI0}) is roughly 20 times larger than a simple estimate (\ref{EBestm1}).
This is because the magnetic field used in eq.(\ref{engBI0}) bends near the inner boundary because of the exclusive type-I condition, and the strength significantly increases.
A larger amount of magnetic energy is sustained in the models with $q>1$.
The energy increases to more than ten times in the case of a strong elastic force ($q>10$).
The energy always increases with $q$ from 
$E_{{\rm mag}} ^{(0)}$ in eq.(\ref{engBI0}), and
the tendency is the same for all models.
The extra component in Model C-I is a toroidal component. Therefore, the toroidal magnetic energy exceeds the poloidal one, which is fixed as $E_{{\rm mag}} ^{(0)}$.
The toroidal field dominated models are possible for large $q$.
The dependence of $l$, a decrease of magnetic energy with $l$, is also common to all models.
This comes from a fact that 
the energy integrated over the crust decreases with a node number $l$,
while the maximum field-strength is almost fixed irrespective of $l$ 
by the maximum strain.

Figure~\ref{Figbx6-I} shows
 the  magnetic  field  and  elastic displacement in
 the $r$-$\theta$ plane for three models with $l=2$ and
 $q=10$.
The crustal region $(r\sin\theta, r\cos\theta)$, ($0.9\le r/R \le 1$) is enlarged by five times in the figure as 
$(\Lambda(r)\sin\theta, \Lambda(r)\cos\theta)$ to display a detailed structure; here,  $\Lambda=1+(R-r)/(2d)$ and $d/R=0.1$.
We use this method throughout to plot functions in a meridional plane\footnote{
A line outside the crust in the figures is a fake caused by graphic software.}.
The poloidal magnetic-function $\Psi$ in Model A-I and Model B-I is described by the combination of the fiducial field of $l=1$ and the extra component of $l=2$; however,
the latter dominates.
In Model C-I in the right panel,
the poloidal field is purely dipolar, and 
additional toroidal-component $B_{\varphi}$ is also shown in color. The angular dependence clearly shows the extra-component with $l=2$.
We show the elastic response for the magnetic field in the lower panels of Fig.~\ref{Figbx6-I}.
The elastic displacement 
${\VEC \xi}_{\rm p}=$ ($\xi_{r}, \xi_{\theta}$)
in a meridional plane is induced by a poloidal magnetic field in Models A-I and B-I.
We use the flow function $X$ defined by eq.(\ref{XibyX.eqn}) to express ${\VEC \xi}_{\rm p}$.
In the representation, the magnitude $|{\VEC \xi}_{\rm p}|$ is expressed by the contour intervals; a stronger $|{\VEC \xi}_{\rm p}|$ for a smaller interval.
The direction is along a constant line of $X$.
For example, the direction is clockwise in a loop with a positive $X$, whereas it is anti-clockwise in a loop with a negative $X$.
The angular distribution of $X$ is approximately described by the multipole with $l=4$ in Model A-I; there are two local minima and two local maxima.
 The distribution in Model B-I (in the middle pannel) is given by $l=3$.
 The angular pattern is produced by a combination of the fiducial field of $l=1$ and the extra component of $l=2$, and the difference originates from their relative weight.
When the extra component dominates, the pattern of $X$ is given by $l=4$, which is similar to that in Model A-I.
The azimuthal component $\xi_{\varphi}$ appears for the mixed poloidal-toroidal magnetic fields as shown in Model C-I (the right panel of Fig.~\ref{Figbx6-I}).
The direction of the elastic displacement, whether $|\xi_{\rm p}| \gg |\xi_{\varphi}|$ or
$|\xi_{\rm p}| \ll |\xi_{\varphi}|$, is approximately  determined by
which component dominates, either the poloidal or the toroidal magnetic field.
In the model shown in the right panel, we find that $|\xi_{\rm p}| \ll |\xi_{\varphi}|$ and  that the toroidal field dominates.
%

\begin{figure}\begin{center}
  \includegraphics[scale=0.85]{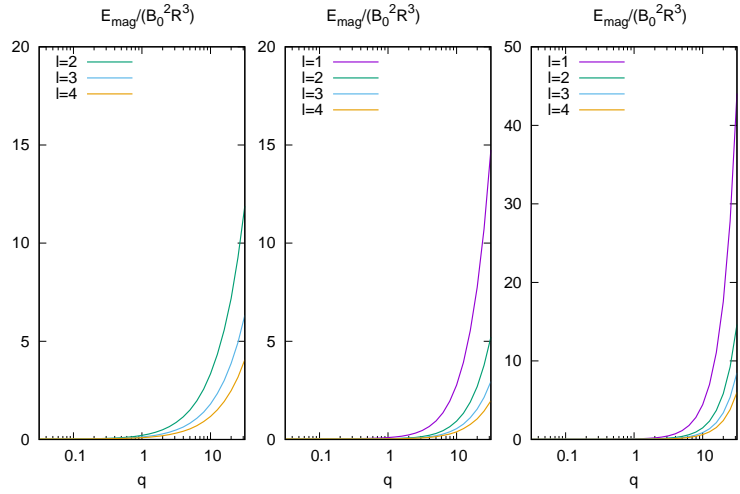}%
\caption{ 
\label{FigE.ACII}
Magnetic energy $E_{\rm mag}/(B_{0}^2R^3)$ in a magneto-elastic equilibrium as a function of $q$.
The left to right panels show results of the extra component of the multi-pole $l$ in Models A-II, B-II, and C-II.
}
\end{center}\end{figure}
\begin{figure}\begin{center}
  \includegraphics[scale=1.0]{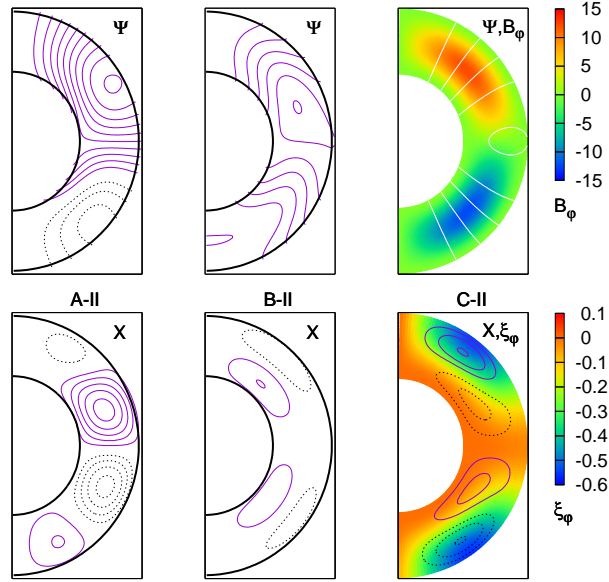}%
\caption{ 
 \label{Figbx6-II}
Magnetic field in top panels and elastic displacement in bottom panels in ``enlarged'' $r$-$\theta$ coordinate for Models A-II, B-II, and C-II with $l=2$ and $q=10$.
The contour intervals of $\Psi$ are $\Delta \Psi/(B_{0}R^2) =0.1$.
The color contour in the right panel represents the toroidal field $B_{\varphi}/B_{0}$.
Contour intervals of $X$ are $\Delta X/(\sigma_{c} d^3) =0.1$
in the left and middle panels, whereas $\Delta X/(\sigma_{c} d^3) =0.01$ in the right panel.
The color contour in the right panel represents the toroidal field $\xi_{\varphi}/(\sigma_{c} d)$.
The dotted lines represent the negative value of the contour level.
}
\end{center}\end{figure}

We examine the effect of the background magnetic field.
The magnetic field in the fiducial model is assumed to penetrate to the core.
We consider Models A-II, B-II, and C-II (see Table~\ref{table1:mylabel}).
The magnetic energy stored in the crust for $q=0$ is given by
 \begin{equation}
 E_{{\rm mag}} ^{(0)}\approx  2.8\times 10^{-2} B_{0}^2R^3
 \approx 4.8  \times 10^{44}
 \left(\frac{B_{0} }{10^{14}{\rm G}}  \right)^{2} {\rm erg}.
   \label{engBII0}
\end{equation}
This value is 0.02 times smaller than that of eq.(\ref{engBI0}) for the same dipolar field strength $B_{0}$.
However, Fig.~\ref{FigE.ACII} shows that the energy increases with $q$, and that
$E_{\rm mag}/(B_{0}^2R^3) >1$ for all models with a sufficiently large value of $q$. 
The magnetic energies are at the same level as those shown in Fig.~\ref{FigE.ACI}.
This means that the extra components dominate, and therefore, the background field is unimportant, whether
the magnetic field is expelled or is penetrating to the core.
%

Figure~\ref{Figbx6-II} shows the magnetic field structure and the elastic displacement for the three models.
The general feature is the same as that considered in Fig.~\ref{Figbx6-I}.
An exception is the pattern of $X$ in Model B-II, wherein the magnitude of $X$ is smaller than that in Model B-I. The angular dependence is given by $l=2$, and there is a node in the radial direction.
%

   \subsection{Energy ratio}

\begin{figure}\begin{center}
    \includegraphics[scale=0.95]{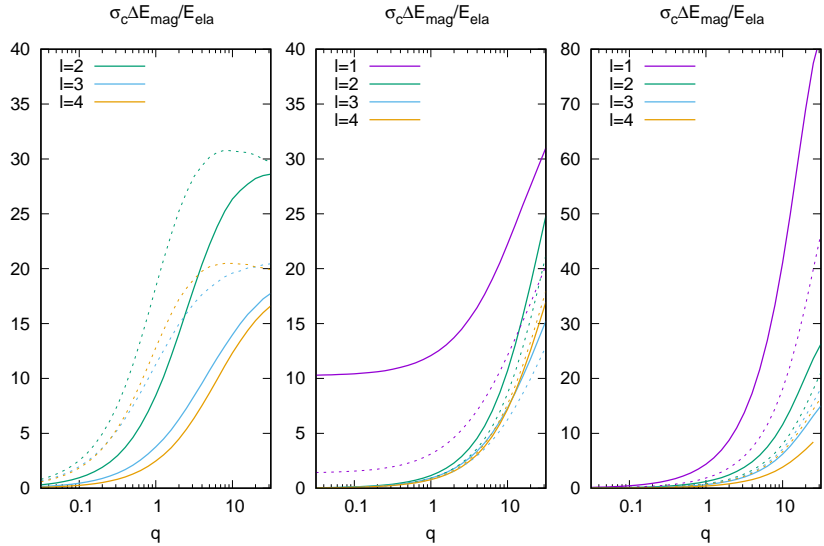}%
\caption{ 
\label{FigMagtoEl}
The ratio of increased magnetic energy 
to elastic energy of the equilibrium models is shown as a function of $q$.
From the left to the right panels, the results of Model A-I, B-I, and C-I are shown by solid lines; those of Models A-II, B-II, and C-II are shown by dotted lines.
}
\end{center}\end{figure}

As shown in Fig.~\ref{FigE.ACI} and Fig.~\ref{FigE.ACII}, the magnetic energy
$E_{\rm mag}$ increases with an elastic strength-parameter $q$.
We compare it with the elastic energy
$E_{\rm ela}$, which is determined by the shear modulus $\mu$ and the maximum shear-strain
 $\sigma_{c}$.
We numerically calculated it, and we found that the energy 
for all models considered in Fig.~\ref{FigE.ACI} and Fig.~\ref{FigE.ACII}
is fitted as
\begin{equation}
 E_{{\rm ela}} \approx ( 0.2 \pm 0.1) \times \mu \sigma_{c}^2R^3 
 \approx 3.5 \times 10^{43}
 \left(\frac{ \mu}{10^{30} {\rm erg~cm^{-3}}}\right)
  \left(\frac{ \sigma_{c}}{10^{-2}}\right)^2 {\rm erg}.
   \label{elsengex}
\end{equation}
This fact means that a ratio 
$\sigma_{c} \Delta E_{\rm mag}/E_{\rm ela} $
increases with $q$, as demonstrated in 
Fig.~\ref{FigMagtoEl} for all models.
 Although there is undetermined parameter  
 $\sigma_{c}$ ($\sigma_{c}<1$ or $\sigma_{c}\ll 1$) involved, we can conclude that
 $E_{\rm mag}>\sigma_{c}^{-1}E_{\rm ela}>E_{\rm ela}$, when the elastic strength $q$ 
 is greater than $q_{c}\sim 5$.
 For a definite discussion, we assume that 
 The material properties  $\mu$ and $\sigma_{c}$
 of the crust are fixed as $\mu=10^{30} {\rm erg~cm^{-3}}$
and  $\sigma_{c}=10^{-2}$.
A magnetar with surface dipole $B_{0}=10^{14}$ G corresponds
to $q=13$.
While the elastic energy 
 $E_{\rm ela}\approx 3.5\times 10^{43}$ erg, 
 the magnetic energy is typically $10^3$ times larger, 
 $E_{\rm mag} \approx 10^{46}$ erg, as shown 
 in Fig.~\ref{FigMagtoEl}.
Besides a large amount of energy, 
a variety of magnetic configurations are allowed.
The constrain however becomes severe with an increase in $B_{0}$.  
We extrapolate the curve of $\sigma_{c} E_{\rm mag}/E_{\rm ela} $
to a much larger $q$.
Assuming a power-law form
$\sigma_{c} E_{\rm mag}/E_{\rm ela} \propto q^{a}$, we numerically determined the index as $a \approx 0.2-0.4$.
We apply it to typical pulsars with $B_{0} \approx 10^{12}$ G, i.e., $q \approx 10^{5}$.
The magnetic energy stored in the crust slightly increases up to
$E_{\rm mag} \approx 5\times 10^{47}$ erg.
%

   \subsection{Model with large toroidal magnetic field}

\begin{figure}\begin{center}
 \includegraphics[scale=0.8]{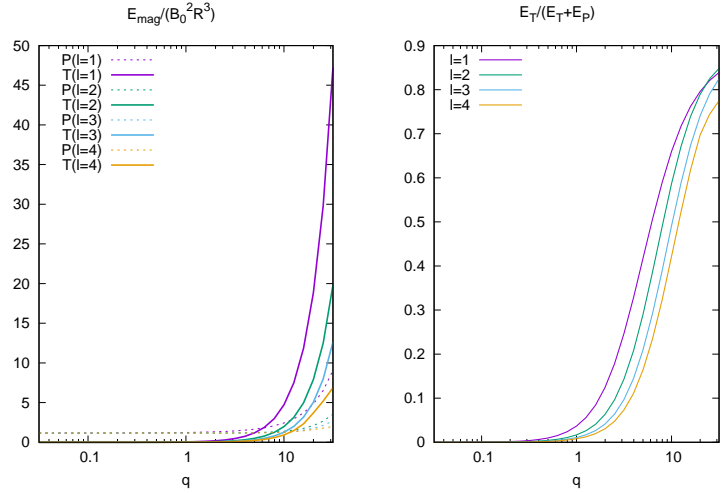}%
\caption{ 
\label{Fig6}
The left panel shows the magnetic energy $E_{\rm mag}/(B_{0}^2R^3)$ as a function of $q$.
The poloidal energy is plotted by a thin line, whereas the toroidal energy is plotted by a thick line.
The right panel shows a fraction of the toroidal energy  $E_{\rm mag,T}/(E_{\rm mag,T}+E_{\rm mag,P})$ as a function of $q$.
}
\end{center}\end{figure}

\begin{figure}\begin{center}
\includegraphics[scale=1.0]{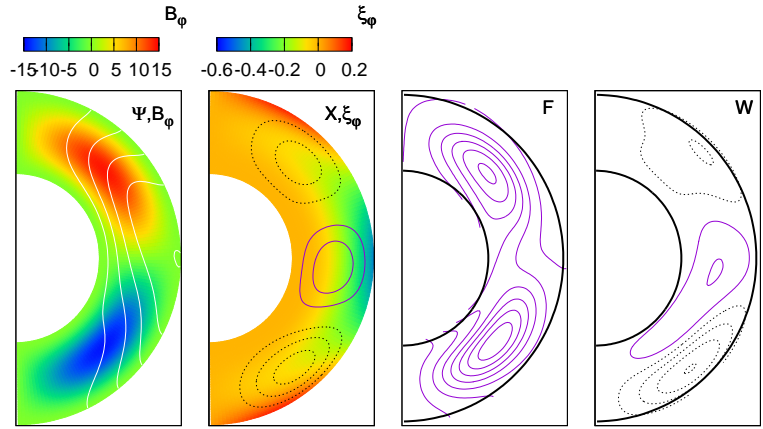}%
\caption{ 
\label{Fig7}
Left panel shows poloidal magnetic-function $\Psi$
by contour intervals of $\Delta \Psi/(B_{0}R^2) =0.1$ and the toroidal field  $B_{\varphi}/B_{0}$ by the color contour.
The second panel shows the contour of $X$
with $\Delta X/(\sigma_{c}d^3)=0.05$
and $\xi_{\varphi}/(\sigma_{c} d)$ by color.
The third and fourth panels are relevant to a decomposition of the Lorentz force.
The third panel shows the contour of $F$ with interval 
$\Delta F/(\mu \sigma_{c}) =2$.
The left panel shows the contour of $W$ with interval 
$\Delta W/(\mu \sigma_{c}) =0.1$.
The dotted lines represent the negative value of the contour level.
}
\end{center}\end{figure}

We consider models with mixed poloidal-toroidal magnetic fields
by simultaneously incorporating the poloidal field $\Psi_{\rm B}$ by eq.(\ref{loopPsi})
and the toroidal field $S_{\rm C}$ by eq.(\ref{loopB3})
to the fiducial field with the type-I boundary condition at $r_c$.
When a condition $S_{\rm C}(\Psi)$ is satisfied, we
have $({\VEC{j}}\times{\VEC{B}})_{\varphi}=0$, and therefore, 
$\xi_{\varphi}=0$.
Our choice is minimizing the displacement, because 
$S_{\rm C}(\Psi)\approx S_{\rm C}(\Psi_{\rm B})$, where 
$\Psi =\Psi_{0}+ \Psi_{\rm B}\approx \Psi_{\rm B}$. 
A variety of models are possible by choices of normalization $N_{{\rm B}}$, $N_{{\rm C}}$.
Here, we fix the ratio as $|N_{{\rm C}}|/|N_{{\rm B}}|=10^2$,
  for which the typical magnetic field strength is $|B_{\varphi}|/|B_{{\rm p}}| \sim 10$.
Thus, the model with the large toroidal magnetic-field is available for large $q$ cases.
The left panel of Fig.~\ref{Fig6} shows the magnetic energy as a function of $q$.
Both the toroidal $E_{\rm mag,T}$ and poloidal $E_{\rm mag,P}$ energies
increase with $q$.
 The right panel of Fig.~\ref{Fig6} shows a fraction of the toroidal energy
 $E_{\rm mag,T} /( E_{\rm mag,T}+E_{\rm mag,P})$.
The limit of $q=0$ is purely poloidal; however, the fraction 
  increases by our choice $|N_{{\rm C}}|/|N_{{\rm B}}|=10^2$.
The toroidal-field dominated model is possible when $q$ is greater than 
$q_{c}\sim 5$.
%

Figure.~\ref{Fig7} shows the mixed poloidal-toroidal magnetic fields with $l=2$ and $q=10$.
The field strength $|B|$ attains the maximum at 
$(r, \theta)=( r_{1}, \pi/4)$ and $(r_{1},3\pi/4)$, where  $r_{1}\approx (r_{c}+R)/2$,
which results from the field confined in the crust.
The elastic displacement induced by the magnetic field is shown in the second panel.
Further, the decomposition of the Lorentz force in the meridian plane is given by
a gradient of $F$ and a curl of 
$A_{\varphi}\equiv\mu W/\varpi$ (see eq.(\ref{decmpFA.eqn})).
The contours of these functions are displayed in the right two panels of Fig.~\ref{Fig7}.
The two peaks of $F$ are located at
$(r, \theta) \approx ( r_{1}, \pi/4)$ and $(r_{1},3\pi/4)$, which are associated with the maximum magnetic pressure imposed extra.
Since the term $F$ may be regarded as a magnetic pressure, a natural normalization may be $B_{0}^2$.
However, we use the same normalization $\mu \sigma_{c}$ to compare $F$ with $W$ in their magnitudes.
The irrotational part by $F$ is approximately 20-50 times larger than the solenoidal part by $W$; it matters with the energy ratio.
The magnetic energy, which is relevant to the irrotational part $F$,
is much larger than the elastic energy.
We do not explicitly solve the perturbation arisen from the irrotational part.
The force associated with $F$
is naturally absorbed by a small change of pressure and gravity.
Here, we considered the force balance in the solenoidal part. 
%

   \subsection{Stress tensor}

\begin{figure}\begin{center}
  \includegraphics[scale=1.0]{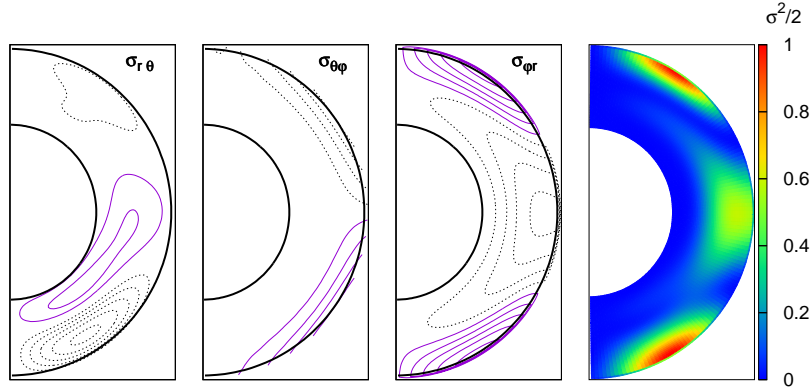}
\caption{ 
\label{Fig8}
Contour of shear stress-tensor $\sigma_{ij}$
normalized by $\sigma_{c}$.
From left to right, three off-diagonal components  $\sigma_{r \theta}/\sigma_{c}$,
 $\sigma_{\theta \varphi}/\sigma_{c}$, 
 $\sigma_{\varphi r}/\sigma_{c}$, and magnitude
 $\sigma_{ij} \sigma^{ij}/(2\sigma_{c} ^2)$ are shown.
Contour interval $\Delta$ is
$\Delta=0.1$ in the range of 
$|\sigma_{r \theta}|/\sigma_{c}\le 0.5$,
$\Delta=0.02$ for 
$|\sigma_{\theta \varphi}|/\sigma_{c}\le 0.08$,
and
$\Delta=0.2$ for 
$|\sigma_{\varphi r}|/\sigma_{c} \le 1$.
The dotted lines represent negative value
$\sigma_{ij}$. 
}
\end{center}\end{figure}

\begin{figure}\begin{center}
  \includegraphics[scale=0.9]{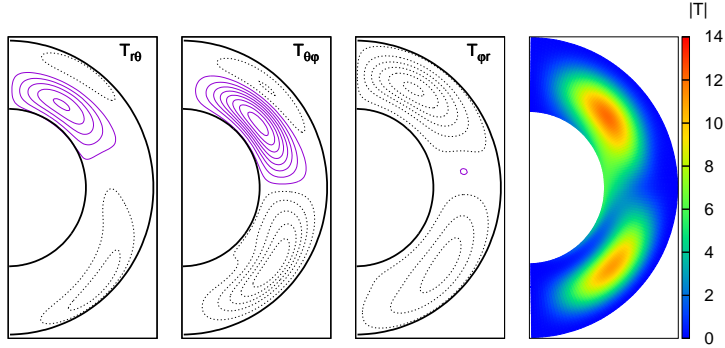}
\caption{ 
\label{Fig9}
The contour of the magnetic stress tensor $T_{ij}$
normalized by $B_{0} ^2$.
From left to right, three off-diagonal components 
 $T_{r \theta}/B_{0} ^{2}$
 $T_{\theta \varphi}/B_{0} ^{2}$, 
 $T_{\varphi r}/B_{0} ^{2}$
 and magnitude
 $|T|\equiv(T_{ij}T^{ij}/2)^{1/2}/B_{0} ^{2}$ are shown.
The contour interval $\Delta$ is
$\Delta=0.1$ for 
$|T_{r \theta}|/B_{0} ^{2}\le 0.6$,
$\Delta=1$ for 
$| T_{\theta \varphi}|/B_{0} ^{2}\le 10$,
and
$\Delta=0.2$ for 
$|T_{\varphi r}|/B_{0} ^{2}\le 1$.
The dotted lines represent the negative value of
$T_{ij}$. 
}
\end{center}\end{figure}

As shown by the results of Models A and B,
the poloidal displacement ${\VEC \xi}_{\rm p}$ is balanced with the force caused by 
the  additional poloidal magnetic-field.
The diagonal components $\sigma_{ii}(i=r, \theta, \varphi)$ and 
$\sigma _{r\theta}$ of the shear stress-tensor
relevant to ${\VEC \xi}_{\rm p}$.
The azimuthal component
 $\xi_{\varphi}$ is induced by the toroidal magnetic field, as shown in Model C.
The components $\sigma _{\varphi r}$ and $\sigma _{\theta \varphi}$ are relevant to $\xi_{\varphi}$.
In Fig.~\ref{Fig8}, we show the three off-diagonal components\footnote{%
It is difficult to display $\sigma_{\varphi r}$ near the surface,
since the contour lines are so crowded
in the figure; however, $\sigma_{\varphi r}=0$ is there.}
for the model considered in Fig.~\ref{Fig7}.
The magnitude $\sigma _{ij}\sigma ^{ij}/2$ of the shear tensor is also shown in the right panel.
We found that
$|\sigma_{\varphi r}|>$ $|\sigma_{\theta \varphi}|>$
$|\sigma _{r \theta}|$ in their magnitudes.
The biggest component $\sigma _{\varphi r}$  almost determines
 $\sigma _{ij}\sigma ^{ij}/2$, because the two right panels of Fig.~\ref{Fig8} are almost the same.
%

In Fig.~\ref{Fig9}, we show the off-diagonal components of magnetic stress
$T_{ij}$(eq.(\ref{Temij.eqn})), and the magnitude
 $(T_{ij}T^{ij}/2)^{1/2}$ for the same model.
We found that
$|T_{\theta \varphi}|>$ $|T_{\varphi r}|>$
$|T _{r \theta}|$ in their magnitudes.
The diagonal components are larger than these off-diagonal components, 
since $|B_{\varphi}|$ $>|B_{\theta}|$ $>|B_{r}|$.
There is no clear relation between
 $T_{ij}$ and $\sigma_{ij}$ concerning the maximum component and the position of the maximum.
The elastic limit is given by 
the Mesis criterion (\ref{criterion}),
which is originally expressed by the shear stress-tensor $\sigma_{ij}$.
An alternative estimator, in which  $\sigma_{ij}$ in the Mesis criterion is replaced by
 $T_{ij}$, is used to find the elastic limit in some papers
\citep[e.g.,][]{2011ApJ...741..123P,2020ApJ...902L..32D}.
There is no linear relationship between 
$\sigma_{ij} $ and $T_{ij}$, and therefore, 
the criterion by $T_{ij}$ is not justified.
The correct relationship between them is 
not $T_{ij}+\mu\sigma_{ij}= 0$ but
$\nabla_{j}(T_{i}^ {j}+\mu\sigma_{i} ^{j})=0$.
%

\section{Concluding remarks}

%
In this paper, we explored an equilibrium of magnetized neutron-star crust with an elastic force.
The Lorentz force produces irrotational acceleration and solenoidal acceleration; the former is balanced with pressure and gravity.
The magnetic force is smaller than these dominant forces and it 
results in a tiny non-spherical deformation.
However, the solenoidal acceleration is not hindered by the dominant forces in the barotropic stars.
We considered the balance with the elastic force in the solid crust, and we demonstrated a large amount of magnetic energy sustained by it.
The magnetic-field configuration is less constrained.
Models with mixed poloidal-toroidal magnetic fields are easily constructed; the 
toroidal magnetic energy may dominate in some models.
The role of elasticity is similar to
that of stratification, in which a buoyancy as a restoring force is weak
but important to stabilize the configuration in fluid stars.
The stable stratification allows an equilibrium model with 
a variety of magnetic-field configuration in nonbarotropic stars.
Further, we found that a large amount of magnetic energy $\sim 10^{47}$ erg
can be stored in a magnetar's crust. 
The magnetic energy $E_{\rm mag}$ exceeds
the elastic energy $E_{\rm ela}$ by a few order of magnitudes.
One might feel something wrong at first; however, they will understand it
by the fact that the solenoidal part of the Lorentz force is much smaller than
the irrotational part in the corresponding model.
The larger irrotational part is associated with larger magnetic energy.
Further, there are also cases in which the magnetic energy 
is not so large; it is comparable to the elastic energy.
The amount of energy depends on the magnetic field configuration.
Our models considered here are not peculiar geometries, and therefore, 
 models with $E_{\rm mag} \gg E_{\rm ela}$ are plausible.
In addition, we showed the spatial distribution of the elastic stress-tensor with the maximum strain. The state corresponds to the onset of the crustal fracture or plastic flow.
These two possibilities are discussed as the transition beyond the critical state, but which case is appropriate is not yet clear at present.
The long-term evolution of the magnetic field
is calculated with a plastic flow
\citep{2020MNRAS.494.3790K,2021MNRAS.502.2097K},
in which the elastic phase is neglected.
In their typical calculations for a magnetar, 
$E_{\rm mag}$ $\sim 10^{45}$ erg
is dissipated in a timescale $\sim 10^{6}$ yr.
The elastic energy is smaller than the Joule energy-loss, and therefore, the effect of the elastic phase is small. 
However, the magnetic field evolution may significantly
be modified, when a burst or flare follows after the elastic limit. 
At the abrupt event, the magnetic field is partially rearranged.
The effect is introduced in the evolution model
\citep[e.g.,][]{2011ApJ...741..123P,2020ApJ...902L..32D}.
The magnetic field around the maximum-stress region is artificially changed
as a quake model, when a criterion reaches a threshold.
Their criterion is based on the magnetic stress tensor; however, it
is irrelevant to the Merier criterion for
the elastic limit, as discussed in Section 3.4.
The stress-tensor of elasticity is different from that of the magnetic field 
in the magnitude and in the spatial distribution.
For example, the position of the maximum amplitude differs
in the two stress tensors. 
The quake in the model is misidentified, and their 
treatment of the magnetic field rearrangement and subsequent evolution are questionable. 
Our model calculated in this paper provides magneto-elastic equilibrium before
the quake.
We can find where the elastic deformation reaches the maximum for the given magnetic field.
The relation is discussed between the magnetic field geometry and burst energy released by rearrangement in the crust-crack models
\citep[e.g.,][]{2019MNRAS.488.5887S,2020ApJ...891...82W,2020ApJ...902L..32D}.
Our model is therefore useful to refine it.
However, our present model contains some approximations.
The shear modulus was treated as a constant; it
generally decreases toward the stellar surface.
We
will find more realistic distribution of shear stress at the critical state by incorporating the shear-modulus varying with density.
The extent of the critical state is related to the magnitude of energy released at the quake.
Further, it is important to know how much energy transfers 
from the available source to the observed bursts or flares
by the rearrangement.
In conclusion, we demonstrated the potential importance of magneto-elastic equilibrium in a magnetized neutron star. Further improvements are therefore necessary.
%

 \section*{Acknowledgements}
%
This work was supported by JSPS KAKENHI Grant Number 
JP17H06361, JP19K03850(YK), JP18H01246, JP19K14712,
JP21H01078(SK), JP20H04728 (KF).
%

 \section*{DATA AVAILABILITY}
%
The data underlying this article will be shared on reasonable request 
to the corresponding author.
%

 \bibliographystyle{mnras}
 \bibliography{kojima21May} 
\end{document}